\documentclass[aps,superscriptaddress,amsmath,amssymb,tightenlines,preprint,11pt]{revtex4-1}
\usepackage{color}
\usepackage[colorlinks=true,linkcolor=blue,citecolor=blue,urlcolor=blue]{hyperref}
\usepackage{comment}
\usepackage{graphicx,dcolumn,braket,bm}
\usepackage[T1]{fontenc} 

\begin{document}

\preprint{RESCEU-18/22}
\preprint{KOBE-COSMO-22-16}

\title{Generation of neutrino dark matter, baryon asymmetry, and radiation after quintessential inflation}

\author{Kohei Fujikura}
	\email{fujikura@penguin.kobe-u.ac.jp}
	\affiliation{Department of Physics, Kobe University, Kobe 657-8501, Japan}
\author{Soichiro Hashiba}
	\email{sou16.hashiba@resceu.s.u-tokyo.ac.jp}
	\affiliation{Research Center for the Early Universe (RESCEU), Graduate School of Science, The University of Tokyo, Tokyo 113-0033, Japan}
	\affiliation{Department of Physics, Graduate School of Science, The University of Tokyo, Tokyo 113-0033, Japan}
\author{Jun'ichi Yokoyama}
	\email{yokoyama@resceu.s.u-tokyo.ac.jp}
	\affiliation{Research Center for the Early Universe (RESCEU), Graduate School of Science, The University of Tokyo, Tokyo 113-0033, Japan}
	\affiliation{Department of Physics, Graduate School of Science, The University of Tokyo, Tokyo 113-0033, Japan}
	\affiliation{Kavli Institute for the Physics and Mathematics of the Universe (Kavli IPMU), WPI, UTIAS, The University of Tokyo, 5-1-5 Kashiwanoha, Kashiwa 277-8583, Japan}
    \affiliation{Trans-Scale Quantum Science Institute, The University of Tokyo, Tokyo 113-0033, Japan}
\date{\today}

\begin{abstract}
We construct a model explaining dark matter, baryon asymmetry and reheating in quintessential inflation model.
Three generations of right-handed neutrinos having hierarchical masses, and the light scalar field leading to self-interaction of active neutrinos are introduced.
The lightest sterile neutrino is a dark matter candidate produced by a Dodelson-Widrow mechanism in the presence of a new light scalar field, while the heaviest and the next heaviest sterile neutrinos produced by gravitational particle production are responsible for the generation of the baryon asymmetry.
Reheating is realized by spinodal instabilities of the Standard Model Higgs field induced by the non-minimal coupling to the scalar curvature, which can solve overproduction of gravitons and curvature perturbation created by the Higgs condensation.
\end{abstract}

\maketitle

\section{introduction}

It is widely believed that the very early stage of the universe experienced exponentially accelerated expansion so-called inflation.
Inflation not only solves fundamental issues such as horizon and flatness problems but also provides seeds of density perturbation. (See e.g.~\cite{Sato:2015dga} for a review of inflation.)
Among many possible variants of inflationary universe models, quintessential inflation~\cite{Peebles:1998qn} is interesting in the sense that the origin of dark energy is attributed to the same scalar field as the inflation-driving field dubbed as the inflaton.  This, however, is not achieved without expenses, as this class of models is associated with a kination or kinetic-energy-dominant regime ~\cite{Spokoiny:1993kt,Joyce:1996cp} after inflation without field oscillation, so that the reheating process after inflation is more involved.  Note that such cosmic evolution is also realized in k-inflation \cite{ArmendarizPicon:1999rj} and a class of (generalized) G-inflation \cite{Kobayashi:2010cm,Kobayashi:2011nu}.

Traditionally, reheating in inflation models followed by a kination regime has been considered by postulating gravitational particle production \cite{Parker:1969au,Zeldovich:1971mw} of a massless minimally coupled scalar field which is produced with the energy density of order of $T_H^4$ at the transition from inflation to kination \cite{Ford:1986sy,Kunimitsu:2012xx}.  Here $T_H=H_{\rm inf}/(2\pi)$ is the Hawking temperature of de Sitter space with the Hubble parameter $H_{\rm inf}$.  In this transition, gravitons are also produced twice as much as the aforementioned scalar field, which acts as dark radiation in the later universe.  Since its energy density relative to radiation is severely constrained by observations of the cosmic microwave background (CMB)~\cite{Planck:2018vyg}, we must assume creation of many degrees of freedom of such massless minimally coupled field whose energy density dissipates in the same way as radiation throughout.  Furthermore, since such a scalar field acquire a large value during inflation due to the accumulation of long-wave quantum fluctuations \cite{Bunch:1978yq,Linde:1982uu,Vilenkin:1982wt}, particles coupled to this field tends to acquire a large mass so that thermalization is not guaranteed. 

In such a situation, two of us \cite{Hashiba:2018iff} calculated gravitational production rate of massive bosons and fermions at the transition from inflation to kination, and concluded that sufficient reheating without graviton overproduction can be achieved if they have an appropriate mass and long enough lifetime, because their relative energy density increases in time with respect to graviton as they redshift in proportion to $a^{-3}(t)$ with $a(t)$ being the cosmic scale factor.  They have further applied the scenario to generations of heavy right-handed Majorana neutrinos to explain origin of radiation, baryon asymmetry, and dark matter in terms of neutrinos \cite{Hashiba:2019mzm}.  

Unfortunately, there are two issues in the previous analysis.  One is that it turned out that in order to explain the full mass spectrum of light neutrinos as inferred by neutrino oscillation, the decay rate of massive right-handed neutrino cannot be small enough to realize appropriate reheating, as shown in Sec.~\ref{sec:reheating}.  The other is the role of the standard Higgs field.  As discussed in \cite{Kunimitsu:2012xx}, if it is minimally coupled to gravity, it suffers from a large quantum fluctuation during inflation~\cite{Starobinsky:1994bd} which will be accumulated to contribute to the energy density of order of $10^{-2}H_{\rm inf}^4$ at the end of inflation. Furthermore its quantum fluctuation is so large that it acts as an unwanted curvaton, which should be removed \cite{Kunimitsu:2012xx}.  

The simplest remedy to the latter problem is to introduce a sufficiently large positive non-minimal coupling to gravity,  so that it has an effective mass squared of $12\xi H_{\rm inf}^2$ during inflation where $\xi>0$ is the coupling constant to the Ricci scalar~\cite{Nakama:2018gll}.  With this coupling, the Higgs field is confined to the origin without suffering from long wave quantum fluctuations.  Furthermore, we can automatically find another source of reheating, namely, the spinodal instability of the Higgs field, as the Ricci scalar will take a negative value in the kination regime and the Higgs field starts to evolve deviating from the origin.  Its subsequent oscillation can create particles of the standard model to reheat the universe as studied in \cite{Nakama:2018gll}.

The purpose of the present paper is to construct a consistent scenario of cosmic evolution generating the observed material ingredients properly in the quintessential inflation model again making use of  three right-handed neutrinos with hierarchical masses inspired by the split seesaw model~\cite{Kusenko:2010ik}. 
The heaviest and the next heaviest right-handed neutrinos realize leptogenesis and explain neutrino oscillation experiments via conventional seesaw mechanism~\cite{Yanagida:1979as,GellMann:1980vs}. The non-minimally coupled SM Higgs field realizes reheating after inflation via spinodal instabilities~\cite{Nakama:2018gll}, while the lightest sterile neutrino and the new light scalar field lead to a successful dark matter production~\cite{DeGouvea:2019wpf}.

In our scenario, baryogenesis through leptogenesis is realized by the decay of the next heaviest right-handed neutrino with mass $M_2$ produced by gravitational particle production.
The heaviest right-handed neutrino is assumed to be much heavier than the Hubble parameter during inflation and only provides the source of CP violation.
We will show the mass range of $M_2$ where the observed amount of the baryon asymmetry is realized.

Finally, the lightest right-handed neutrino can constitute cold dark matter if it is nearly stable so that its life-time is longer than the age of the universe.
The simplest production mechanism of such a light right-handed neutrino dark matter is through neutrino oscillations between the left-handed and right-handed neutrinos known as Dodelson-Widrow mechanism~\cite{Dodelson:1993je}, apart from gravitational particle production introducing a non-minimal coupling as assumed in \cite{Hashiba:2019mzm}.
However, constrains from X-ray observation~\cite{Boyarsky:2005us,Boyarsky:2006fg,Boyarsky:2006ag,Boyarsky:2007ay} combined with constraints from phase space analysis~\cite{Tremaine:1979we,Boyarsky:2008ju,Gorbunov:2008ka} and Lyman-$\alpha$ forest~\cite{Viel:2005qj,Boyarsky:2009ix,Yeche:2017upn,Palanque-Delabrouille:2019iyz} excludes this simplest possibility.
(See e.g. Refs.~\cite{Drewes:2016upu,Boyarsky:2018tvu} for review of the sterile neutrino dark matter.)

Successful production mechanisms of the right-handed neutrino dark matter such as a resonant production~\cite{Shi:1998km} and production with new physics in addition to the right-handed neutrinos~\cite{Shaposhnikov:2006xi,Khalil:2008kp,Kaneta:2016vkq,Biswas:2016bfo,Seto:2020udg,DeRomeri:2020wng,Belanger:2021slj,Cho:2021yxk} have been suggested.
Among them, we focus on the possibility of the sterile neutrino dark matter production with a secret active neutrino self-interaction originally proposed in Ref.~\cite{DeGouvea:2019wpf}.
In this scenario, a new light complex singlet scalar field which induces a self-interaction of active neutrino is introduced.
Production rate of the active neutrino in the early universe is enhanced by the new interaction, and the resultant relic density of the lightest sterile neutrino can make up all of the dark matter in the parameter space consistent with current constraints.
We will calculate the relic abundance of the lightest sterile neutrino dark matter by analytically solving the Boltzmann equation under some reasonable approximations and show that keV-scale sterile neutrino can explain relic dark matter density when a mass scale of the new light scalar field is around MeV scale.

The rest of the paper is organized as follows.
In Sec.~\ref{sec:right-handed neutrino}, we review basic features of right-handed neutrinos.
In Sec.~\ref{sec:reheating}, we see that reheating of the universe by the decay of gravitationally produced right-handed neutrino cannot be achieved, but the non-minimal coupling between the SM Higgs and the scalar curvature can lead to the efficient reheating.
In Sec.~\ref{sec:baryogenesis}, we explain baryogenesis through leptogenesis by gravitationally produced sterile neutrino.
Then, we analytically calculate the relic density produced by the lightest right-handed neutrino with a secret self-interaction of the left-handed neutrinn in Sec.~\ref{sec:dark matter}.
Sec.~\ref{sec:conclusion} is devoted to the conclusion.

\section{Hierarchical sterile neutrinos}\label{sec:right-handed neutrino}

In this section, we review general features of the right-handed neutrinos.
We consider the following Lagrangian density for the right-handed neutrino $\nu_{Ri}$ ($i=1,2,3$) with hierarchical masses $M_i$ ($M_1\ll M_2 \ll M_3$)
\begin{align}
-\mathcal{L}_N = h_{\alpha i} \bar{L}_\alpha \tilde{H}_{\rm SM}N_i +\dfrac{1}{2}M_i \bar{\nu}^c_{Ri} \nu_{Ri}+{\rm h.c.} \,.\label{eq: Yukawa}
\end{align}
In this expression, $L_{\alpha}=(\nu_{L\alpha},e_{L\alpha}
)^{\rm T},~H_{\rm SM}$ and $h_{\alpha i}$ are the SM lepton doublet, the SM Higgs doublet and the Yukawa coupling constants, respectively.
The suffices $\alpha=e,\mu,\tau$ denote the generation of the SM leptons and $\psi^c$ denotes the charge conjugation of the $\psi$ field.
After the electroweak symmetry breaking, the SM Higgs field acquires the vacuum expectation value, $\langle H_{\rm SM}\rangle =v_{\rm SM}/\sqrt{2}$ where $v_{\rm SM}\simeq 246\,{\rm GeV}$, leading to the Dirac mass terms.
The mass matrix of neutrinos is then given by
\begin{align}
		-\mathcal{L}_{\rm mass} = \frac{1}{2} (\bar{\nu}_{L},\bar{\nu}_{R}^c)\mathcal{M} 
	\begin{pmatrix}
		\nu^c_{L}\\
		\nu_{R}
	\end{pmatrix}+~{\rm h.c.}\,,~
	\mathcal{M}=
	\begin{pmatrix}
	0&m_D\\
	m_D^{\rm T} & D_M
    \end{pmatrix},
\end{align}
where $(m_D)_{\alpha i} \equiv h_{\alpha i}v_{\rm SM}/\sqrt{2}$ is the $3\times 3$ Dirac mass matrix, and $D_M \equiv {\rm diag}(M_1,M_2,M_3)$, respectively.
The matrix, $\mathcal{M}$, can be diagonalized by the unitary matrix $U$:
\begin{align}
U^{\dag} \mathcal{M}U^*	= {\rm diag}(m_{\nu_{\alpha'}},m_{N_{I}}),~(\alpha'=1,2,3,~I=1,2,3).
\end{align}

Assuming $m_D \ll M_i$, at the leading order, the unitary matrix can be expressed as~\cite{Grimus:2000vj}
\begin{align}
	U=
	\begin{pmatrix}
		U_{\rm PMNS}&\theta\\
		-\theta^\dag U_{\rm PMNS}&{\bm 1}_{3\times 3}
	\end{pmatrix},~\theta \equiv m_D D_M^{-1} \,.
\end{align}
In this expression, $U_{\rm PMNS}$ is the Pontecorvo-Maki-Nakagawa-Sakata (PMNS) matrix~\cite{Maki:1962mu} defined by the following relation:
\begin{align}
	U_{\rm PMNS}^{\dag} M_\nu U_{\rm PMNS}^* = {\rm diag}(m_{\nu_{1}},m_{\nu_{2}},m_{\nu_{3}}),~M_\nu\equiv -m_DD_M^{-1}m_D^{\rm T}. \label{eq:active neutrino mass}
\end{align}
For $\theta_{\alpha i}\ll 1$, mass eigenstates $\nu_{\alpha'}$ and $N_{i}$ called active and sterile neutrinos are explicitly given by
\begin{align}
	\nu_{L\alpha} = (U_{\rm PMNS})_{\alpha \alpha'}\nu_{\alpha'}+\theta_{\alpha i}N_i^c,~\nu^c_{Ri}= -\theta^*_{\alpha i}(U_{\rm PMNS})_{\alpha \alpha'} \nu_{\alpha'} + N^c_i .
\end{align}
One can see from the above expression that the active neutrinos, $\nu_{\alpha'}$, and the sterile neutrinos, $N_I$, almost correspond to a linear combination of $\nu_{L\alpha}$ and $\nu_{Ri}$ itself, respectively.
Also, the mass of the sterile neutrino $N_i$ is almost identical to the Majorana mass of the right-handed neutrino, $m_{N_I} \simeq \delta_{Ii} M_i$ for $\theta_{\alpha i}\ll 1$, and hence, we do not distinguish them in what follows.

There are several constraints on active neutrino masses from observations of neutrino oscillations such as the Super-Kamiokande~\cite{Fukuda:1998mi}, KamLAND~\cite{Araki:2004mb} and the MINOS~\cite{Adamson:2011ig}.
Absolute values of active neutrino mass-squared differences are constrained as $m_{\rm sol}^2\equiv |m_{\nu_2}^2-m_{\nu_1}|^2=7.59\times 10^{-5}\,{\rm eV}^2$ and $m_{\rm atm}^2\equiv|m_{\nu_3}^2-m_{\nu_1}^2|=2.32\times 10^{-3}\,{\rm eV}^2$.
One cannot take $h_{\alpha i}$ arbitrarily free since it is related to active neutrino masses through the relation Eq.~\eqref{eq:active neutrino mass}.
To make the lightest sterile neutrino dark matter, it will turn out in Sec.~\ref{sec:dark matter} that Yukawa coupling of the lightest sterile neutrino becomes vanishingly small, $\sum_\alpha |\tilde{h}_{\alpha 1}|^2\ll 1$ where $\tilde{h}\equiv U^\dagger_{\rm PMNS}h$.
Resultant contributions to $m_{\nu_{2,3}}$ from $\tilde{h}_{\alpha 1}$ are negligible amount and are decoupled in the seesaw formula.
Under this setup with assuming normal mass hierarchy $m_{\nu_3}> m_{\nu_2}>m_{\nu_1}$, constraints on active neutrino masses are simplified as
\begin{align}
 m_{\nu_3}\simeq 0.05\,{\rm eV} ~{\rm and}~ m_{\nu_2}\simeq 0.01\,{\rm eV}. \label{eq:neutrino oscillation}
\end{align}
The above condition will be used in Sec.~\ref{sec:reheating} and Sec.~\ref{sec:baryogenesis}.

\section{Reheating}\label{sec:reheating}

In this section, we discuss reheating in our model.
Before discussing the reheating mechanism in detail, we would like to clarify our setup.
We consider a spatially flat Friedmann-Lema\'itre-Robertson-Walker (FLRW)
background, $\mathrm{d}s^2=-\mathrm{d}t^2+a^2(t)\mathrm{d}{\bm x}^2$, where $a(t)$ denotes the scale factor.
We consider following smooth transition from the de Sitter phase to the kination phase~\cite{Hashiba:2018iff}:
\begin{align}
	a^2 (\eta) = \frac{1}{2}\left[ \left(1-\tanh \frac{\eta}{\Delta \eta}\right)\frac{1}{1+H_{\rm inf}^2\eta^2}+\left(1+\tanh\frac{\eta}{\Delta \eta} \right)(1+H_{\rm inf}\eta)\right].  \label{eq:scale factor}
\end{align}
In this expression, $H_{\rm inf}$ is the Hubble parameter during inflation, $\eta$ is the conformal time which satisfies $\mathrm{d}t=a(\eta)\mathrm{d}\eta$ and $\Delta \eta>0$ parametrizes the time scale of the transition which depends on the Lagrangian of quintessential inflation, k-inflation, or G-inflation.
For $\Delta \eta \lesssim 1.7 H_{\rm inf}$, $a^2 (\eta)$ is monotonically increasing function and remains positive for all $\eta$.
With this parametrization, the scale factor is normalized to unity around the end of inflation, so that $\Delta\eta$ is identical to the physical time scale of transition, $\Delta t$ practically.

The energy density of particles created gravitationally during this transition
has been calculated in the literature~\cite{Kunimitsu:2012xx,Hashiba:2018iff,Hashiba:2018tbu} as
\begin{equation}
    \rho_\varphi= C_{\varphi}H_{\rm inf}^4a^{-4}(t),~(C_\varphi \simeq 5\times 10^{-3},~\Delta t=H_{\rm inf}^{-1}), \label{eq:massless minimally coupled scalar energy density}
\end{equation}
for a massless minimally coupled scalar field $\varphi$, and
\begin{equation}
     \rho_{\rm mf}=C_{\rm mf}m^2H_{\rm inf}^2 e^{-4m\Delta t}a^{-3}(t),~(C_{\rm mf}\simeq 2\times 10^{-3}), \label{eq:fermion energy density}
\end{equation}
for a massive fermion with mass $m$.
In these expressions, $C_{\varphi}$ and $C_{\rm mf}$ are numerically determined for the scale factor given by Eq.~\eqref{eq:scale factor}.

Spinodal instability of the non-minimally coupled Higgs field, on the other hand, sets in after inflation, and it takes some time until its energy starts to dissipate when its energy density behaves as \cite{Nakama:2018gll}
\begin{align}
	\rho_{\rm Higgs}= C_{\rm Higgs} H_{\rm inf}^4 \dfrac{a^4 (t_s)}{a^4(t)}. \label{eq:energy density of spinodal instability}
\end{align}
Here $a(t_{\rm s})$ is the scale factor at the moment when growth of spinodal instability terminates.
Typically, $a^3(t_{\rm s})\sim \mathcal{O}(10)$, but its precise value depends on $\xi$ and $\Delta t$ where $\xi$ is the size of non-minimal coupling of the SM Higgs with gravity. (See appendix. \ref{appendix:spinodal instability} for the precise definition of $\xi$.)
$C_{\rm Higgs}$ is sensitive to the value of $\xi$, which is order of $C_{\rm Higgs}\simeq  0.05\sim \mathcal{O}(1)$ for $\xi \sim \mathcal{O}(1)$.

One can investigate the parameter region where the decay products of the sterile neutrino are the dominant source of radiation by comparing
the energy density (\ref{eq:energy density of spinodal instability}) with that of heavy neutrinos given by (\ref{eq:fermion energy density}) at their decay time, $t_{\rm d}$.
It defined by the equality $H(t_{\rm d})=\Gamma_{N_i}$, where  $\Gamma_i =\gamma_i M_i$ with $\gamma_i \equiv \sum_\alpha |\tilde{h}_{\alpha  i}|^2/(8\pi)$~\cite{Fukugita:1986hr,Covi:1996wh} being the tree-level decay rate of $N_i$ in the rest frame.
We assume that the decay takes place during kination.
In the kination dominated era, the Hubble parameter is given by $H(t)=H_{\rm inf}/a^3(t)$, and then, the moment when the decay takes place denoted by $t_{\rm d}$ is determined by the condition $\Gamma_i=H_{\rm inf}/a^3(t_{\rm d})$.
$N_i$ decays into SM particles that behave as radiation, and hence, $\rho_{N_i}(t>t_{\rm d})\propto a^{-4}$ where $\rho_{N_i}$ is the energy density of $N_i$.
By comparing the radiation energy density produced by spinodal instability of the SM Higgs field Eq.~\eqref{eq:energy density of spinodal instability} at $t=t_{\rm d}$, it turns out that the dominant component of radiation is sourced by $\rho_{N_i}$ when
\begin{align}
	\dfrac{C_{\rm mf}}{a^4(t_{\rm s})C_{\rm Higgs}} \left(\dfrac{M_i}{H_{\rm inf}}\right)^{\frac{5}{3}} e^{-4M_i \Delta t} \gamma_i^{-1/3}>1,
\end{align}
is satisfied.
The left-hand side of the above equation takes the maximum value when $M_i/H_{\rm inf}\simeq 0.42$ for $\Delta t =H_{\rm inf}^{-1}$.
Therefore the above condition can be expressed as 
\begin{align}
	\gamma_i  < 8.5 \times 10^{-5} \times \left(\dfrac{C_{\rm mf}}{a^4(t_{\rm s})C_{\rm Higgs}}\right)^3. \label{eq:reheating condition}
\end{align}

As we explained in the previous section, the Yukawa coupling $\tilde{h}_{\alpha i}$ must be chosen in such a way that Eq.~\eqref{eq:neutrino oscillation} is satisfied to explain neutrino oscillation experiments.
This implies that there exists non-trivial bound on $\gamma_i$.
Neglecting contributions from $\tilde{h}_{\alpha 1}$, Eq.~\eqref{eq:active neutrino mass} can be rewritten as the following explicit expresisons:
\begin{align}
     -\dfrac{v_{\rm SM}^2}{2M_2}\tilde{h}_{22}^2-\dfrac{v_{\rm SM}^2}{2M_3}\tilde{h}_{23}^2=m_{\nu_2}\,,\nonumber\\
     -\dfrac{v_{\rm SM}^2}{2M_2}\tilde{h}_{32}^2-\dfrac{v_{\rm SM}^2}{2M_3}\tilde{h}_{33}^2=m_{\nu_3}\,,\label{eq:mass matrix}\\
     -\dfrac{v_{\rm SM}^2}{2M_2}\tilde{h}_{22}\tilde{h}_{32}-\dfrac{v_{\rm SM}^2}{2M_3}\tilde{h}_{23}\tilde{h}_{33}=0\,.\nonumber
\end{align}
The last term comes from the off-diagonal term of Eq.~\eqref{eq:active neutrino mass}.
For normal mass hierarchy, $m_{\nu_\alpha}$ are given by Eq.~\eqref{eq:neutrino oscillation}.
Then the magnitude of the Yukawa coupling squared is bounded as
\begin{align}
     \sum_\alpha |\tilde{h}_{\alpha 3}|^2 
     &\simeq   |\tilde{h}^2_{23}| + |\tilde{h}_{33}^2|\nonumber\\
     &=\left|\dfrac{2M_3m_{\nu_2}}{v_{\rm SM}^2}+\dfrac{m_{\nu_2}}{m_{\nu_3}}\tilde{h}_{33}^2\right| + |\tilde{h}_{33}^2| \nonumber\\
     &\gtrsim \dfrac{2M_3 m_{\nu_2}}{v_{\rm SM}^2}. \label{eq:lower bound Yukawa}
\end{align}
In the second equality, we used Eq.~\eqref{eq:mass matrix}.
A lower bound on $\sum_\alpha |\tilde{h}_{\alpha 2}|^2$ is obtained in a similar manner.
The resultant bounds are summarized as $\sum_\alpha |\tilde{h}_{\alpha i}|^2\gtrsim 2M_i m_{\nu_2}/v_{\rm SM}^2$, for $i=2,3$ in the case of normal hierarchy, and consequently, $\gamma_i$ is bounded below.
Using the bound on $\gamma_i$~\eqref{eq:reheating condition}, we find that radiation is dominated by the decay of $N_i$ when 
\begin{align}
	M_i \lesssim 5.1\times 10^7\,{\rm GeV}\times \left(\frac{0.1}{a^4(t_{\rm s})C_{\rm Higss}}\right)^3, \label{eq:upper bound}
	\end{align}
is satisfied.
In the opposite region, radiation is mainly sourced by the spinodal instability of the SM Higgs field.

\begin{figure}[t]
\begin{center}
\includegraphics[clip, width=7.5cm]{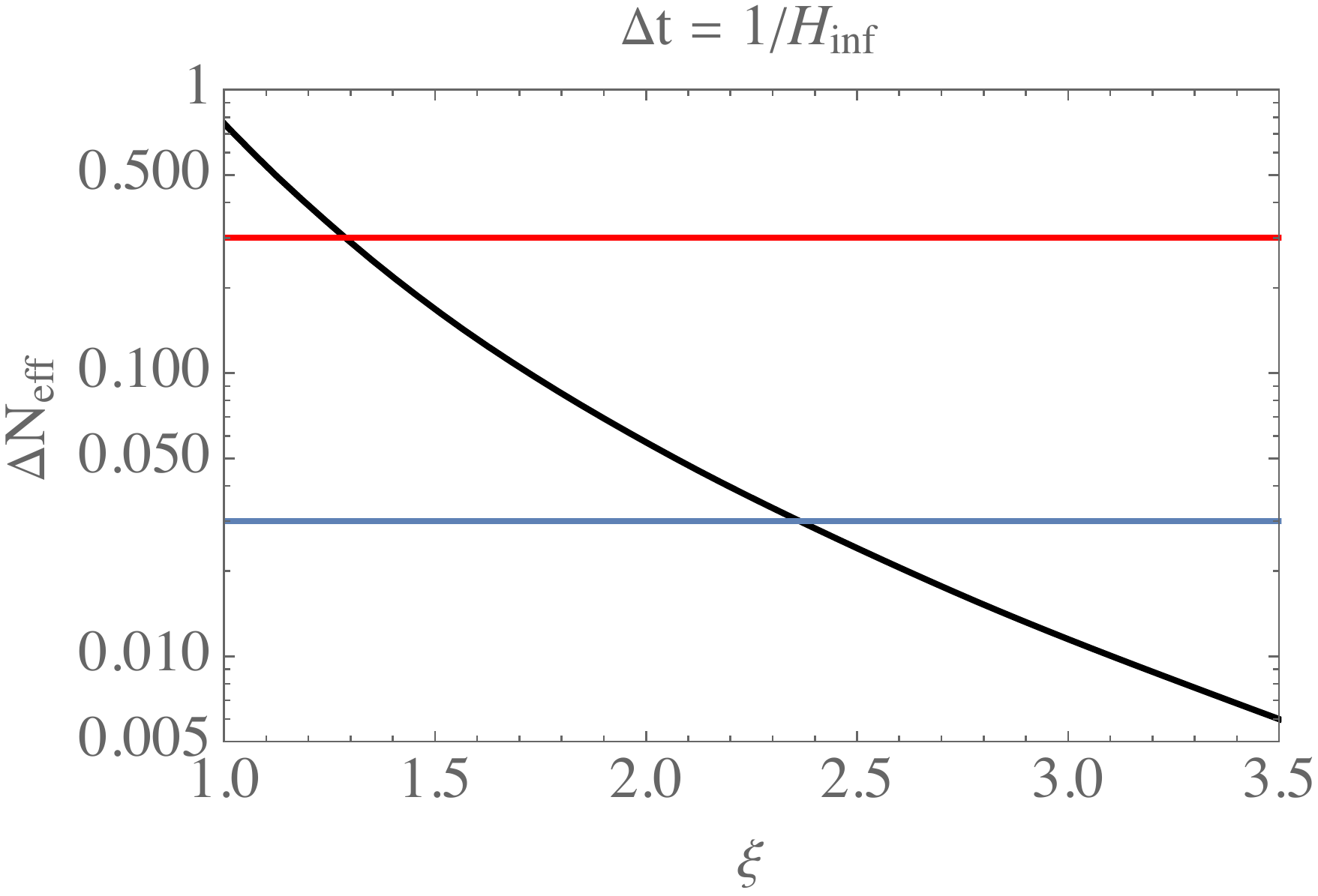}
\end{center}
\caption{
 The dependence of $N_{\rm eff}$ at the photon decoupling on the Higgs non-minimal coupling, $\xi$, for $\Delta t = H_{\rm inf}^{-1}$ with the quartic coupling of the SM Higgs field, $\lambda_{\rm SM}=0.01$, is shown. The red contour represents the current limit $\Delta N_{\rm eff} =0.3$ of the Planck data~\cite{Planck:2018vyg}, while the blue contour corresponds to $\Delta N_{\rm eff}=0.03$ which is the futuristic target sensitivity of the next generation stage-4 ground-based cosmic microwave background experiment~\cite{CMB-S4:2016ple,Abazajian:2019eic}.
\label{fig:Neff}
}
\end{figure}

The above condition can be further translated into the upper bound on the reheating temperature in the case reheating is realized by the decay products of massive neutrinos rather than spinodal instability.
The time when reheating takes place denoted by $t_{\rm RH}$, is defined by the condition, $\rho_{\rm kin}(t_{\rm RH})=\rho_{\rm rad}(t_{\rm RH})$ where $\rho_{\rm kin}$ is the kinetic energy density of the inflaton.
Since $\rho_{\rm kin}\propto a^{-6}$, one can parameterize the kinetic energy of inflaton as $\rho_{\rm kin}(t)=3M_{\rm Pl}^2H_{\rm inf}^2 /a^6(t)$, where $M_{\rm Pl}\simeq 2.44\times 10^{18}\,{\rm GeV}$ is the reduced Planck mass.
Then, the reheating temperature $T_{\rm RH}$ can be expressed by
\begin{align}	
	T_{\rm RH} &= \left(\dfrac{10}{3\pi^2 g_*(T_{\rm RH}) }\right)^{\frac{1}{4}}C_{\rm mf}^\frac{3}{4}\gamma_i^{-\frac{1}{4}} \left(\dfrac{M_i}{H_{\rm inf}}\right)^{\frac{5}{4}}\dfrac{H_{\rm inf}^2}{M_{\rm Pl}}\,e^{-3M_i\Delta t}\\
	&\lesssim 2.2\times 10^{-3}\dfrac{M_i^2}{M_{\rm Pl}}\gamma_i^{-1/4}
\end{align}
Here, $g_*$ is the number of effective degrees of freedom of the primordial hot plasma.
In this calculation, we used $\Delta t = H_{\rm inf}^{-1},~g_*(T_{\rm RH})=10.75$ and $M_i/ H_{\rm inf}=0.42$, which gives the maximum reheating temperature.
Again using the lower bound on $\gamma_i$ given by~\eqref{eq:lower bound Yukawa}, we obtain
\begin{align}
	T_{\rm RH}\lesssim 0.46\,{\rm MeV}\times \left(\dfrac{0.1}{a^4(t_{\rm s})C_{\rm Higgs}}\right)^{\frac{21}{4}}. \label{eq:reheating bound}
\end{align}
The above bound on the reheating temperature may be marginally consistent with the success of the big bang nuclesynthesis (BBN), $T_{\rm RH}\gtrsim 0.5\,{\rm MeV}$~\cite{Kawasaki:1999na} when $a^4(t_s)C_{\rm Higgs}$ is small.
Numerical calculations, however, reveal that $a^4(t_{\rm s})C_{\rm Higgs}$ is larger than $0.1$ unless $\xi$ is smaller than or very close to unity. (See Fig.~\ref{fig:Neff} and Eq~\eqref{eq:C Higgs}.)
Thus, for $\xi \gtrsim \mathcal{O}(1)$, spinodal instability of the SM Higgs field should be the most dominant source of radiation to realize the efficient reheating, and we focus on this parameter space in the following, which corresponds to the case the decaying sterile neutrino masses are naturally heavier than the threshold \eqref{eq:upper bound} and much higher reheating temperature is realized by spinodal instability.

Before evaluating the reheating temperature in this case, let us 
discuss the issue of the dark radiation problem caused by gravitational particle production of gravitons.
Gravitationally created gravitons are never in thermal equilibrium with the SM particle, and thus, they become a problematic source of dark radiation.
The energy density of the total relativistic components, $\rho_{\rm tot}$, can be decomposed as the energy desnities of thermal plasma consisting of the SM particles, $\rho_{\rm rad}$, and gravitons:
\begin{align}
	\rho_{\rm tot}= \rho_{\rm rad}+\rho_{\rm GW}=\frac{\pi^2}{30}g_*(T) T^4+\rho_{\rm GW}.
\end{align}
Here, $\rho_{\rm GW}$ is the energy density of the produced gravitons, respectively.
Since the graviton satisfies the same equation of motion as a massless minimally coupled scalar field, its energy density is twice as much as that of the minimally coupled scalar field, $\rho_{\rm GW}=2\rho_{\varphi}$.
A constraint on dark radiation components is conventionally given in terms of the effective number of neutrinos species at the photon decoupling parameterized by
\begin{align}
	\rho_{\rm tot}(t_{\rm ph}) = \rho_\gamma (t_{\rm ph}) \left[1+\frac{7}{8}\left(\dfrac{4}{11}\right)^{\frac{4}{3}}N_{\rm eff}\right].
\end{align}
In this expression, $\rho_\gamma (t_{\rm ph})$ is the energy density of the photon at the photon decoupling, $t=t_{\rm ph}$.
Using entropy conservation, the deviation from the standard value of $N_{\rm eff}$ at the photon decoupling can then be parameterized as 
\begin{align}
    \mathrm{\Delta}N_{\rm eff}  &= \frac{4}{7}\left(\frac{4}{11}\right)^{-\frac{4}{3}}g_*(t_{\rm ph})\left(\frac{g_* (t_{\rm ph})}{g_*(t_{\rm th})}\right)^{\frac{1}{3}}\left.\left(\dfrac{\rho_{\rm GW}}{\rho_{\rm rad}}\right)\right|_{t=t_{\rm th}}, \label{eq:Delta N}
\end{align}
where $t_{\rm th}$ is the moment when thermalization of SM fields takes place.
Assuming thermalization takes place sufficiently early, $g_*(t_{\rm th})=106.75$~\cite{Husdal:2016haj}, and using $g_*(t_{\rm ph})\simeq 3.38$ which is the value including an effect of non-fully decoupled neutrinos at the electron-positron annihilation~\cite{Mangano:2005cc,Akita:2020szl}, the expression Eq.~\eqref{eq:Delta N} can be rewritten as follows.
\begin{align}
\mathrm{\Delta}N_{\rm eff}\simeq 2.36\times \left.\left(\dfrac{\rho_{\rm GW}}{\rho_{\rm rad}}\right)\right|_{t=t_{\rm th}}. \label{eq:Delta N eff}
\end{align}
The recent Planck data~\cite{Planck:2018vyg} gives the constraint $\mathnormal{\Delta}N_{\rm eff}<0.30$ at $95\%$ confidence level.

We numerically evaluate the energy density created by spinodal instability of the SM Higgs field.
The detailed numerical procedure is summarized in Ref.~\cite{Nakama:2018gll}. (See also appendix.~\ref{appendix:spinodal instability}.)
The dependence of $\Delta N_{\rm eff}$ on the non-minimal coupling $\xi$ is shown in Fig.~\ref{fig:Neff}.
There we use the SM Higgs quartic coupling $\lambda_{\rm SM}=0.01$ as a reference value although its precise value depends on the scale of $H_{\rm inf}$ and physics around this scale because of the renormalization group effect.
In the figure, we take $\Delta t = H_{\rm inf}^{-1}$, but we confirm that $\mathnormal{\Delta}N_{\rm eff}$ is less sensitive to $\Delta t$ since a smaller $\Delta t$ makes both spinodal instability and graviton production large.
Also, $\mathrm{\Delta}N_{\rm eff}$ is less sensitive to tiny quartic coupling $\lambda_{\rm SM}\lesssim 0.01$.
On the other hand, from this figure, one finds that $\mathnormal{\Delta}N_{\rm eff}$ is strongly sensitive to the value of $\xi$.
This simply reflects the fact that a large $\xi$ makes spinodal instability large, and thus, the energy density of $\phi$ is amplified.
These results are in agreement with the earlier work~\cite{Nakama:2018gll}.
The current constraint $\mathrm{\Delta}N_{\rm eff}<0.3$ is compatible with $\xi \gtrsim 1.4$.
The future observation by the next generation stage-4 ground-based cosmic microwave background experiment will probe the parameter space $\mathrm{\Delta}N_{\rm eff}>0.03$~\cite{CMB-S4:2016ple,Abazajian:2019eic} corresponding to $\xi \lesssim 2.3$.

Since both the energy density of the Higgs field (\ref{eq:energy density of spinodal instability}) and that of gravitons have only a weak dependence, from Eq.~\eqref{eq:Delta N eff} we can find a one-to-one correspondence between
$\Delta N_{\rm eff}$ and $C_{\rm Higgs}$ which is basically determined by $\xi$ as
\begin{align}
	C_{\rm Higgs}\simeq 7.9\times 10^{-2}\times \left(\dfrac{0.30}{\mathnormal{\Delta}N_{\rm eff}}\right)\times \left(\frac{1}{a^4(t_{\rm s})}\right),~\left(\Delta t =H^{-1}_{\rm inf}\right). \label{eq:C Higgs}
\end{align}
The scale factor at $t=t_{\rm RH}$ is denoted by $a_{\rm RH}$ and reheating temperature $T_{\rm RH}$ can be evaluated as
\begin{align}
	a_{\rm RH}=\sqrt{\dfrac{3}{C_{\rm Higgs}}}\dfrac{M_{\rm Pl}}{H_{\rm inf}}a^{-2}(t_{\rm s}),~T_{\rm RH} = \left(\frac{10}{3}\frac{C_{\rm Higgs}^3 a^{12}(t_{\rm s})}{\pi^2g_*(T_R)}\right)^{\frac{1}{4}} \dfrac{H_{\rm inf}^2}{M_{\rm Pl}}. \label{eq:reheating temperature spinodal}
\end{align}
%
For $H_{\rm inf}=10^{13}\,{\rm GeV},~g_*(T_{\rm RH})=106.75$ and $\mathnormal{\Delta}N_{\rm eff}= 0.3$, one obtains $T_R\simeq 1.4\times 10^6\,{\rm GeV}$ leading to the efficient reheating compared to Eq.~\eqref{eq:reheating bound}.

Before closing this section, let us estimate the thermalization temperature during kination denoted by $T_{\rm th}$, which can be much higher than $T_{\rm RH}$ in general.
Since the oscillating SM Higgs field after growth of spinodal instabilities mainly decay into the ${\rm SU(2)_{\rm W}}$ gauge bosons, thermalization of SM fields is expected to take place via scattering of the SM fields mediated by ${\rm SU(2)_{\rm W}}$ gauge interaction.
Hence we determine the thermalization temperature by the condition $\Gamma_{\rm SU(2)_W}(T_{\rm th})=H_{\rm kin}(T_{\rm th})$ where $\Gamma_{\rm SU(2)_W}$ and $H_{\rm kin}$ are the reaction rate of scattering of SM fields through ${\rm SU(2)_{\rm W}}$ gauge interaction and the Hubble parameter during kination, respectively.
The reaction rate is roughly given by $\Gamma_{\rm SU(2)_{W}}\sim \alpha_W^2T$ by dimensional analysis, where $\alpha_W\equiv g_W^2/(4\pi)$ with $g_W$ being ${\rm SU(2)_W}$ gauge coupling.
$H_{\rm kin}$ can be expressed in terms of temperature in the following way.
It follows from $\rho_{\rm kin}\propto a^{-6}$ and $\rho_{\rm rad}(t_{\rm RH})=\rho_{\rm kin}(t_{\rm RH})$ that $\rho_{\rm kin}(t)=\rho_{\rm rad}(t_{\rm RH})a^6_{\rm RH}/a^6(t)$.
Assuming there is no entropy generation other than spinodal instability of the SM Higgs field, which is indeed the case in our scenario, entropy conservation leads to
\begin{align}
	\rho_{\rm kin}(T)= \frac{\pi^2}{30}\dfrac{g^2_*(T)}{g_*(T_{\rm RH})}\frac{T^6}{T_{\rm RH}^2}.
\end{align}
The Hubble parameter during kination is given by $H^2_{\rm kin}(T)=\rho_{\rm kin}(T)/(3M_{\rm Pl}^2)$.
Thus, the condition $\Gamma_{\rm SU(2)_W}(T_{\rm th})=H_{\rm kin}(T_{\rm th})$ yields the following relation.
\begin{align}
    T_{\rm th}\simeq 8.4\times 10^{-3}\sqrt{T_{\rm RH}M_{\rm Pl}}.
\end{align}
Here, we put $g_W= 0.6$ and $g_*(T_{\rm th})=g_*(T_{\rm RH})=106.75$.
Using Eqs.~\eqref{eq:C Higgs} and \eqref{eq:reheating temperature spinodal}, $T_{\rm th}$ can be further rewritten as 
\begin{align}
    T_{\rm th}\simeq 1.6\times 10^{-3} H_{\rm inf} \left(\dfrac{0.30}{\mathrm{\Delta}N_{\rm eff}}\right)^{\frac{3}{8}}. \label{eq:thermalization temperature}
\end{align}
As is obvious from the above expression, $T_{\rm th}$ is much higher than $T_{\rm RH}$.
This expression will be used in the next section.

\section{Baryogenesis}\label{sec:baryogenesis}

In this section, we discuss baryogenesis through leptogenesis in this model.
We focus on the scenario where $N_2$ is abundantly produced by gravitational particle production and its decay is responsible for the generation of lepton asymmetry.
(See, however, Ref.~\cite{Chun:2007np} for the possibility of the thermal leptogenesis on kination background.)
The presence of $N_3$ provides CP-violation to $N_2$ decay via one-loop effects.
The out-of-equilibrium condition can be satisfied if $N_{2,3}$ are never in thermal equilibrium.

For $M_3 \gg H_{\rm inf}\gtrsim M_2$, the energy density of $N_3$ field is exponentially suppressed and is negligible compared to that of $N_2$.
In this parameter range, lepton asymmetry is created dominantly by the decay of $N_2$ with the abundance
\begin{align}
	n_L = \epsilon_2 \frac{\rho_{N_2}}{M_2}.
\end{align}
In this expression, $\rho_{N_{2}}$ is the energy density of $N_{2}$ field produced by gravitational particle production.
$\epsilon_{2}$ is the asymmetry parameter, which parameterizes the CP violation of $N_2$ decay defined by
\begin{align}
	\epsilon_2 \equiv \frac{\Gamma(N_2 \to L +H_{\rm SM}) -\Gamma(N_2\to \bar{L}+\bar{H}_{\rm SM})}{\Gamma(N_2 \to L +H_{\rm SM}) +\Gamma(N_2\to \bar{L}+\bar{H}_{\rm SM})}.
\end{align}
Here, $\Gamma (X)$ is the decay rate of the reaction $X$ and final states are summed over all lepton flavors, and the charged and neutral components of the lepton doublet and SM Higgs field.
Non-zero $\epsilon_i$ is provided by interference between decay amplitudes of tree-level and one-loop diagrams.
One-loop corrections from vertex and self-energy~\cite{Flanz:1994yx,Covi:1996wh,Buchmuller:1997yu} lead to
\begin{align}
\epsilon_2 = \frac{1}{8\pi} \frac{\sum_{j \neq 2} {\rm Im}\left[ \{ (h^\dagger h)_{2j} \}^2 \right]}{(h^\dagger h)_{22}} \left\{ f^V \left( \dfrac{M_j^2}{M_2^2} \right) + f^M \left( \dfrac{M_j^2}{M_2^2} \right) \right\},
\end{align}
where 
\begin{align}
f^V(x) = \sqrt{x} \left[ -1 + (x + 1) \ln \left( \frac{x + 1}{x} \right) \right],~
	f^M(x) =  \frac{\sqrt{x}}{x-1}.
\end{align}
For hierarchical mass $M_1/M_2 \ll M_2/M_3 <1$, $\epsilon_2$ is approximately given by
\begin{align}
		\epsilon_2 &\simeq \dfrac{3}{16\pi} \cfrac{1}{(h^\dagger h)_{22}} {\rm Im}\left[ \{ (h^\dagger h)_{23} \}^2 \right] \cfrac{M_2}{M_3}.
\end{align}
Using ${\rm Im}\,(h^\dagger h)_{22}=0$ and the definition of active neutrino mass Eq.~\eqref{eq:active neutrino mass}, one can obtain an upper bound on $\epsilon_2$~\cite{DiBari:2005st,Vives:2005ra} for normal hierarchy:
\begin{align}
	|\epsilon_2| \lesssim  \frac{3}{8\pi}\dfrac{M_2m_{\nu_3}}{v_{\rm SM}^2}. \label{eq:efficiency of CP violation}
\end{align}
For hierarchical active neutrino masses, $\epsilon_2$ takes almost maximum value unless CP-violating phase is accidentally suppressed.
We will see that this upper bound leads to bounds on $M_2$ in the following analysis.

The total lepton number density at reheating temperature is estimated from Eq.~\eqref{eq:fermion energy density} as 
\begin{align}
\left.	n_L\right|_{T=T_{\rm RH}} =  \epsilon_2 C_{\rm mf}\left(\frac{C_{\rm Higgs}}{3}\right)^{3/2}M_2 \frac{H_{\rm inf}^5}{M_{\rm Pl}^3} a^6(t_{\rm s})e^{-4M_2\Delta t} .
\end{align}
The cosmic entropy at reheating temperature can be also calculated from Eq.~\eqref{eq:reheating temperature spinodal} and is given by 
\begin{align}
	s(T_{\rm RH})= \dfrac{2}{45}\left(\dfrac{10}{3}\right)^{\frac{3}{4}}\left(\pi^2 g_*(T_{\rm RH})\right)^{\frac{1}{4}}C_{\rm Higgs}^\frac{9}{4}\dfrac{H_{\rm inf}^6}{M_{\rm Pl}^3}a^9(t_{\rm s}).
\end{align}
In this calculation, we approximate the effective number of relativistic degrees of freedom of entropy as that of the energy deensity.
The generated lepton asymmetry is then converted into baryon asymmetry via sphaleron process:
\begin{align}
	\frac{n_B}{s} = C \left.\frac{n_L}{s}\right|_{T=T_{\rm RH}}, \label{BL}
\end{align}
Here, $C =-12/37$ in the SM with crossover electroweak transition~\cite{Khlebnikov:1988sr,Harvey:1990qw}.
Then, baryon-to-entropy ratio is expressed as
\begin{equation}
\begin{split}
	\left|\dfrac{n_B}{s}\right|&\simeq 
	1.3\times10^{-3}\times  \epsilon_2 \left(\dfrac{\mathrm{\Delta}N_{\rm eff}}{0.30}\right)^\frac{3}{4}\left(\dfrac{g_*(T_{\rm RH})}{106.75}\right)^{-\frac{1}{4}}\dfrac{M_2}{H_{\rm inf}}e^{-4M_2\Delta t}\\
	&\lesssim 1.3\times 10^{-7}\times\left(\frac{M_2}{10^{13}\,{\rm GeV}}\right)\left(\dfrac{\mathrm{\Delta}N_{\rm eff}}{0.30}\right)^\frac{3}{4}\left(\dfrac{g_*(T_{\rm RH})}{106.75}\right)^{-\frac{1}{4}}\dfrac{M_2}{H_{\rm inf}}e^{-4M_2\Delta t}.
	\label{eq:generated baryon asymmetry}
	\end{split}
\end{equation}
In the second line, we have used the inequality \eqref{eq:efficiency of CP violation}.

Since the factor $M_2/H_{\rm inf}e^{-4M_2\Delta t}$ takes the maximum value $9.2\times 10^{-2}$ for $M_2/H_{\rm inf}=1/4$ when $\Delta t=H_{\rm inf}^{-1}$, the generated baryon asymmetry is bounded above.
The observed baryon-to-entropy ratio is given by $n^{\rm obs}_B/s\simeq  8.59\times 10^{-11}$~\cite{Planck:2018vyg}.
Thus, to produce the observed baryon asymmetry, $M_2$ is restricted as 
\begin{align}
	M_2 \gtrsim 7.3\times 10^{9}\,{\rm GeV}\times \left(\dfrac{0.30}{\mathrm{\Delta}N_{\rm eff}}\right)^{\frac{3}{4}}.\label{eq:lower bound}
\end{align}
The equality is satisfied when $\epsilon_2$ takes its maximum value and $M_2/H_{\rm inf}=1/4$ is realized.

On the one hand, the Hubble parameter during inflation is constrained by a measurement of tensor mode for CMB anisotropy as $H_{\rm inf}\lesssim 4.6\times 10^{13}\,{\rm GeV}$~\cite{BICEP:2021xfz}.
The generated baryon asymmetry is thus bounded from above as
\begin{align}
	\left|\dfrac{n_B}{s}\right|\lesssim 5.9\times 10^{-7}\times\left(\dfrac{\mathrm{\Delta}N_{\rm eff}}{0.30}\right)^\frac{3}{4}\left(\dfrac{g_*(T_{\rm RH})}{106.75}\right)^{-\frac{1}{4}} \left(\dfrac{M_2}{H_{\rm inf}}\right)^2e^{-4M_2\Delta t}.
\end{align}
The equality is satisfied when $H_{\rm inf}=4.6\times 10^{13}\,{\rm GeV}$.
Assuming $\Delta t = H_{\rm inf}^{-1}$ and using the observed amount of the baryon asymmetry, the above inequality can be translated into the bound on $M_2/H_{\rm inf}$ for $\Delta t = H_{\rm inf}^{-1}$.
This condition turns out to be $M_2 \lesssim 1.2\times 10^{14}\,{\rm GeV}$ for $g_*(T_{\rm RH})=106.75$ and $\Delta N_{\rm eff} =0.3$.
Combining with the lower bound on $M_2$ Eq.~\eqref{eq:lower bound}, we find that successful production of the observed baryon asymmetry can be realized for the following window:
\begin{align}
	7.3\times 10^{9}\,{\rm GeV}\times \left(\dfrac{0.30}{\mathrm{\Delta}N_{\rm eff}}\right)^{\frac{3}{4}} \lesssim M_2\lesssim  1.2\times 10^{14}\,{\rm GeV}.
\end{align}

Assuming $\epsilon_2$ takes the maximum value Eq.~\eqref{eq:efficiency of CP violation}, for $M_2=7.3\times 10^{9}\,{\rm GeV},~H_{\rm inf}=4M_2\simeq 2.9\times 10^{10}\,{\rm GeV}$ and $\mathrm{\Delta}N_{\rm eff}=0.30$, the observed baryon asymmetry can be achieved.
In this benchmark point, the reheating temperature becomes $T_{\rm RH}\simeq 12\,{\rm GeV}$.
As is obvious from the expression Eq.~\eqref{eq:thermalization temperature}, the out-of-equilibrium condition is maintained for this benchmark point.
We will show that $N_1$ make up all dark matter in this benchmark point in the next section.

\section{Dark Matter}\label{sec:dark matter}

In this section, we discuss dark matter production in our model.
We consider the scenario where the lightest sterile neutrino $N_1$ is a dark matter candidate, whose mass is around ${\rm keV}$ scale.
$N_1$ must be stable so that its signal must be below the level detectable by the previous and ongoing X-ray observations such as Chandra, XMM-Newton, Suzaku and NuSTAR~\cite{Perez:2016tcq,Kopp:2021jlk,Abazajian:2017tcc,Roach:2022lgo}.
This imposes stringent constraints on mixing angle $\sin^2 2\theta_{\alpha 1}\lesssim \mathcal{O}(10^{-14})$ for $M_1\sim 50\,{\rm keV}$.
For such a tiny mixing angle, contributions to active neutrino masses from $h_{\alpha 1}$ can be neglected, and thus, there is no constraint on $h_{\alpha 1}$ from the neutrino oscillation experiments.

Let us consider the scenario proposed in Ref.~\cite{DeGouvea:2019wpf} in which a self-interaction between left-handed neutrino is introduced.
The Lagrangian density is given by
\begin{align}
	-\mathcal{L}_\nu \supset m_{\Phi}^2|\Phi|^2+\dfrac{\lambda_\Phi}{2} \Phi \nu_{L a} \nu_{L a} +{\rm h.c.}. \label{eq:neutrino secret self-inetraction}
\end{align}
Here, $\Phi$ is a light SM gauge singlet complex scalar field, which possesses $B-L$ charge $-2$.
The suffix $a=e,\mu$ or $\tau$ represents single flavor of the left-handed neutrino.
An extension of multi-flavor interactions is straightforward and we consider a single flavor for an illustrative purpose.
The above interaction is not invariant under SM gauge symmetry and is regarded as an effective interaction after the electroweak symmetry breaking.
Several models that produce the effective interaction Eq.~\eqref{eq:neutrino secret self-inetraction} have been proposed~\cite{Berryman:2018ogk,Blinov:2019gcj,Dev:2021axj}.
In these models, $B-L$ is regarded as a good symmetry and we assume that $B-L$ symmetry is only violated by the Majorana mass term in Eq.~\eqref{eq: Yukawa}.
Under this assumption, the new interaction Eq.~\eqref{eq:neutrino secret self-inetraction} does not wash out the generated lepton asymmetry.
A secret self-interaction of the sterile neutrinos, $y_{ij}\nu^c_{Ri} \nu^c_{Rj} \Phi^*$, may exist, but we also assume tiny Yukawa coupling $y_{ij}\ll h_{\alpha i}$ and has no effects on leptogenesis discussed in the previous section.

$N_1$ is never in thermal equilibrium with the primordial hot plasma due to the smallness of the mixing angle and is non-thermally produced by the neutrino oscillation between $\nu_{R1}\simeq N_1$ and $\nu_{L \alpha}$ as in the original model~\cite{Dodelson:1993je}.
In our scenario, the lepton asymmetry is of same order magnitude of the baryon asymmetry, and hence, the lepton chemical potential is negligibly small.
For the light sterile neutrino mass $M_1 \ll H_{\rm inf}$, initial abundance of $N_1$ produced by gravitational particle production Eq.~\eqref{eq:fermion energy density} is also negligible as here we do not assume non-minimal coupling to gravity realized in Ref.~\cite{Hashiba:2019mzm}.
Under this setup, time evolution of the phase-space distribution function of $N_1$, with fixed ratio of active neutrino energy to the cosmic temperature $x\equiv E/T$, is governed by the Boltzmann equation~\cite{Dodelson:1993je,Abazajian:2001nj}:
\begin{align}
\frac{\mathrm{d}f_{N_1}(x,z)}{\mathrm{d}z} = \dfrac{\Gamma \sin^2 2\theta_{\rm eff}}{4H(z)z}f_{\nu_{L a}}(x). \label{eq:Boltzmann equation}
\end{align}
In this expression, $z\equiv 1/T,~\Gamma$ and $f_{\nu_{La}}(x)$ are the inverse cosmic temperature, the interaction rate and the Fermi-Dirac distribution function for $\nu_{La}$, respectively.
$\sin^2 2\theta_{\rm eff}$ is the effective active-sterile mixing angle including finite-temperature effect given by
\begin{align}
	\sin^2 2\theta_{\rm eff} \equiv \dfrac{\Delta^2 \sin^2 2\theta_{a1}}{\Delta^2\sin^2 2\theta_{a1} +\Gamma^2/4+(\Delta \cos2\theta_{a1}-V^T)^2}. \label{eq:effective mixing angle}
\end{align}
Here, $\Delta \simeq  M_1^2/(2E)$ and $V^T$ is the thermal potential of $\nu_a$, respectively.
The interaction rate and the thermal potential can be schematically decomposed as $\Gamma=\Gamma_{\rm SM}+\Gamma_{\Phi}$ and $V^T= V^T_{\rm SM}+V^T_{\Phi}$, where $\Gamma_{\rm SM},~V^T_{\rm SM}$ and $\Gamma_{\Phi},~V^T_{\Phi}$ denote the SM contributions and contributions from $\Phi$ field, respectively.
The SM contributions are given by $\Gamma_{\rm SM}\simeq G_F^2E T^4$~\cite{Abazajian:2001nj} and $V^T_{\rm SM}\simeq G_F ET^4/M_W^2 $~\cite{Notzold:1987ik,Quimbay:1995jn}, where $G_F$ and $M_W$ are Fermi constant and $W$ gauge boson mass, respectively.

We restrict ourselves to consider the parameter regime where the main production is by on-shell exchange of $\Phi$ for simplicity.
(However, off-shell production can also produce correct relic dark matter density if $\lambda_\Phi$ is sufficiently large~\cite{DeGouvea:2019wpf}.)
Using narrow-resonance approximation, on-shell contribution to $\Gamma_\Phi$ can be estimated as~\cite{DeGouvea:2019wpf}
\begin{align} 
\Gamma_{\Phi}= \dfrac{\lambda_{\Phi}^2m_{\Phi}^2T}{8\pi E^2}\log\left(1+e^{-w/x}\right),~w\equiv m_\Phi^2 z^2/4\label{eq:on-shell}
\end{align}
For $w/x\gtrsim1$, the interaction rate is exponentially suppressed.
Therefore, the efficient on-shell production of $\Phi$ is possible only for $m_\Phi\lesssim T$.
In this parameter regime, the thermal potential from $\Phi$ are approximately given by~\cite{DeGouvea:2019wpf}
\begin{align} 
V^T_{\Phi}= \lambda_\Phi^2 T^2/(16E) \label{eq:thermal potential}
\end{align}
For $\Delta > V^T,\,\Gamma$, the effective mixing angle can be approximated by the vacuum angle $\sin^2 2\theta_{\rm eff}\simeq \sin^2 2\theta_{a1}$ as can be seen from the expression Eq.~\eqref{eq:effective mixing angle}.

When the effective mixing angle can be approximated by the vacuum angle, the Boltzmann equation Eq.~\eqref{eq:Boltzmann equation} can be expressed by the formal integrated expression:
\begin{align}
	\dfrac{f_{N_1}}{f_{\nu_{L a}}} = \dfrac{\lambda_\Phi^2}{8\pi}\dfrac{M^*_{\rm Pl}}{m_\Phi}\frac{1}{x^2}\sin^2 2\theta_{a1}\int_{w_i}^{w_f} \mathrm{d} w \,G(w,x),~G(w,x)\equiv w^{1/2} \log \left(1+e^{-w/x}\right). \label{eq:integrated Boltzmann equation}
\end{align}
In this calculation, we change the integration variable from $z$ to $w$ and use the Hubble parameter during the radiation domination era, $H(z)=1/(M^{*}_{\rm Pl}z^2)$ with $M^*_{\rm Pl}\equiv \sqrt{90/(\pi^2 g_*(z))}M_{\rm Pl}$.
$w_i$ and $w_f$ parameterize the integration range and $g_*(T)$ is assumed to be constant during this interval.

For $x\sim 1$, where the highest population of $\nu_{La}$ is realized, the integrand behaves as $G(w)\propto \sqrt{w}$ for $w\ll1$ and $G(w)\propto \sqrt{w}e^{-w}$ for $w\gg 1$, and has a peak around $w\simeq 1$.
This implies that $N_1$ is dominantly produced at $w\simeq 1$, while it is suppressed for both small and large $w$.
Therefore, approximations that are used to derive Eq.~\eqref{eq:integrated Boltzmann equation} should be justified only for $w\simeq 1$ corresponding to the temperature $T\sim m_{\Phi}$.
Hence the approximation $\sin^2 2\theta_{\rm eff}\simeq \sin^2 2\theta$ needs to be justified around $T\sim m_{\Phi}$.
This requirement becomes $\Delta > V^T,\Gamma$ at $T\sim m_\Phi$ leading to conditions $m_\Phi < M_1/\lambda_\Phi$ and $m_\Phi < \mathcal{O}(100)\,{\rm MeV}(M_1/\,{\rm keV})^{1/3}$.
In this light mass regime, one may approximately set the integration range as $w_i=0$ and $w_f=\infty$ since contributions from $w\ll 1$ and $w\gg 1$ in the integrand of \eqref{eq:integrated Boltzmann equation} is suppressed.
Note that we slightly overestimate relic energy density of $N_1$ under this approxiamtion.

By setting $w_i=0$ and $w_f=\infty$, $f_{N_1}$ can be approximately calculated by doing integration of $g(w,x)$ in Eq.~\eqref{eq:integrated Boltzmann equation}: 
 \begin{align}
 f_{N_1} \simeq  A_1\dfrac{\lambda_\Phi^2}{8\pi}\dfrac{M^*_{\rm Pl}}{m_\Phi}\sin^2 2\theta_{a1} \sqrt{\dfrac{1}{x}}\,f_{\nu_{L_a}}.\label{eq:phase-space distribution}
 \end{align}
 Here, $A_1\simeq 0.769$ is the numerical constant arising from the integration of $G(w)$.
 It should be emphasized that the above phase-space distribution is colder than the one in the original DW mechanism as noticed in the original paper~\cite{DeGouvea:2019wpf}. 
The relic number density of $N_1$ can be estimated by integrating Eq.~\eqref{eq:phase-space distribution} and is given by
\begin{align}
	\dfrac{n_{N_1}}{n_{\nu_{L a}}} =A_1\dfrac{A_2}{A_3}\dfrac{\lambda_{\Phi}^2}{8\pi}\frac{M_{\rm Pl}^*}{m_{\Phi}}\sin^2 2 \theta_{a1},
\end{align}
where $A_2\simeq 1.15,A_3\simeq 1.80$ are numerical constants and $n_{\nu_{L a}}$ is the number density of $\nu_{L a}$.
Using the number density of $\nu_{L a}$ at the present time $n_{\nu_{La}} \simeq 112\,{\rm cm}^{-3}$ and the value of the critical density $\rho_0 \simeq 1.05\times 10^{-5}h^{-2}\,{\rm GeV}/{\rm cm^3}$, the relic energy density of the sterile neutrino dark matter turns out to be
\begin{align}
	\Omega_{N_1}h^2\simeq 0.12\times \left(\dfrac{M_1}{50\,{\rm keV}}\right)\times \left(\dfrac{\sin^2 2\theta_{a1}}{10^{-14}}\right)\times \left(\dfrac{100\,{\rm MeV}}{m_\Phi}\right) \times \left(\dfrac{\lambda_\Phi}{2.3 \times 10^{-4}}\right)^2 .\label{eq:dark matter density}
\end{align}
 We confirm that this analytic result is in good agreement with original results for $m_\Phi < M_1/\lambda_\Phi$ where the above expression is applicable.
 In our parameter choice, the coupling $\lambda_\Phi$ is too small to be probed by laboratory experiments~\cite{Berryman:2018ogk}, but there exists relevant cosmological and astrophysical constraints.
$m_{\Phi}$ must be heavier than the few MeV in order to not spoil the BBN~\cite{Blinov:2019gcj}.
In addition, the light scalar mediator which couples to the active neutrino causes distinct changes in the supernovae collapse dynamics and gives the stringent constraint on the light mass regime~\cite{1987ApJ...322..795F,1988ApJ...332..826F,Chang:2022aas,Chen:2022kal}.

For the benchmark point $M_1\simeq 50\,{\rm keV},~\sin^2 2\theta =10^{-14},~m_{\Phi}=100\,{\rm MeV}$ and $\lambda_{\Phi}=2.3\times 10^{-4}$, constraints mentioned above combined with X-ray observations are marginally satisfied and produce the correct dark matter relic density.

Finally, we comment on the constraint on the free-streaming length of the produced sterile neutrino dark matter and the validity of approximations used to derive the relic dark matter density.
When the dark matter is produced with high velocity dispersion, it erases the small scale fluctuation and conflicts with the observation such as Lyman-$\alpha$ forest~\cite{Palanque-Delabrouille:2019iyz,Dekker:2021scf}.
Since this constraint depends on the phase-space distribution of the produced dark matter Eq.~\eqref{eq:phase-space distribution}, which is different from the one in the original Dodelson-Widrow mechanism, to obtain a precise constraint we need the model-dependent analysis, which is beyond scope of the present paper.
In our analysis, we have neglected off-shell contributions.
Since off-shell contributions to $\Gamma_\Phi$ and $V^T_{\Phi}$ are proportional to $\lambda_\Phi^4$, an extra suppression factor $\lambda_\Phi^2$ appears compared to on-shell contributions.
For $\lambda_\Phi \ll 1$, off-shell contribution is subdominant, and hence, we can safely neglect this contribution.
Also, usage of the Hubble parameter during the radiation domination era is justified for $T_{\rm RH}\gg m_\Phi$.
This condition is satisfied for the benchmark point taken above.

\section{conclusion}\label{sec:conclusion}

In this paper, we have constructed a phenomenologically viable model which explains production of radiation, baryon asymmetry and dark matter within the framework of quintessential inflation model which also accommodates dark energy by construction.
Three right-handed neutrinos with hierarchical masses and a new light singlet scalar field are introduced.
We have shown that the reheating achieved by the decay of the heavy sterile neutrinos lead to the reheating temperature as low as MeV scale contrary to Ref.~\cite{Hashiba:2019mzm} because of the constraint from the neutrino oscillation experiments if the lightest sterile neutrino is the dark matter candidate.
In our scenario, reheating is achieved by the SM Higgs spinodal instability triggered by the non-minimal coupling of the scalar curvature.
Since gravitationally produced gravitons become a problematic dark radiation, the energy density of the SM Higgs field must be larger than that of gravitons to avoid the dark radiation problem, which gives a lower bound on the size of non-minimal coupling. (See Fig.~\ref{fig:Neff}.)
After the end of inflation, the next heaviest sterile neutrino are abundantly produced by gravitational particle production and its decay is responsible for the leptogenesis.
The heaviest right-handed neutrino is assumed to be heavier than the Hubble parameter during the inflation, and it only provides a source of CP violation for the decay of the next heaviest sterile neutrino.
To realize this scenario, the mass of the next heaviest right-handed neutrino is restricted.
The lightest right-handed neutrino with keV scale mass is a dark matter candidate introducing a new light scalar field whose mass scale is around of MeV scale.
A secret self-interaction of the active neutrino induced by the new light scalar field enhances the production rate of the lightest neutrino and the lightest sterile neutrino can explain correct dark matter relic density~\cite{DeGouvea:2019wpf}.
We analytically calculated the relic density of the lightest sterile neutrino under reasonable approximations for parameter space where the production is dominated by the on-shell exchanging of the new light scalar field.

\begin{acknowledgments}  KF acknowledges useful comments of Kohei Kamada and Tomohiro Nakama. K.F. is supported by JSPS Grant-in-Aid for Research Fellows Grant No. 22J00345. SH was supported by the Advanced Leading Graduate Course for Photon Science (ALPS). The work of JY was supported by JSPS KAKENHI, Grant JP15H02082 and Grant on Innovative Areas JP15H05888.
\end{acknowledgments}

\appendix

\section{Reheating by spinodal instabilities of the SM Higgs field}\label{appendix:spinodal instability}

In this appendix, we explain the reheating scenario proposed in Ref.~\cite{Nakama:2018gll} where the energy density of the SM Higgs field from spinodal instability is a main source of cosmic entropy.
The Lagrangian density of the SM Higgs field with a non-minimal coupling to the Ricci scalar is given by
\begin{align}
	\mathcal{L}_\phi= \sqrt{-g}\left( -\frac{1}{2}\partial_\mu \phi \partial^\mu \phi-\frac{1}{2}m^2\phi^2 -\frac{\lambda_{\rm SM}}{4} \phi^4 -\frac{1}{2}\xi R \phi^2\right).
\end{align}
Here, $\phi$ represents the real and neutral component of the SM Higgs field.
Couplings with other SM particles are omitted.
The SM Higgs quartic coupling $\lambda_{\rm SM}$ is assumed to be positive up to the scale of inflation, $H_{\rm inf}$.
Variational principle leads to the equation of motion for $\phi$:
\begin{align}
	\ddot{\phi}+3H\dot{\phi}-\frac{\nabla^2}{a^2}\phi  +m^2\phi+\xi R\phi+\lambda_{\rm SM}\phi^3 =0. \label{eq:equation of motion}
\end{align}
Here, $\dot{\phi}$ is the derivative with respect to the cosmic time $t$.
Since the mass of $\phi$ field is presumably much smaller than the scale of inflation and reheating when $H_{\rm inf}\gg m$, we neglect it in the following analysis.

Let us briefly explain the reheating mechanism.
During inflation, $R\cong 12H_{\rm inf}^2>0$, and hence, the non-minimal coupling becomes effective positive mass term of $\phi$ field.
Therefore $\phi$ is trapped at the origin initially.
As the kination regime commences, $R$ becomes negative, which drives $\phi$ to the minimum $\phi_M = \sqrt{-\xi R/\lambda_{\rm SM}}$.
During this process, super-horizon modes of $\phi$ grow due to spinodal instability, and consequently, the energy density of $\phi$ is amplified.
Since the scalar curvature behaves as $R\propto a^{-6}(\eta)$ during kination, spinodal instability is soon shut off and oscillation of $\phi$ takes place with the quartic potential.
Finally, oscillating $\phi$ field mainly decay into SM gauge bosons~\cite{Figueroa:2015rqa,Enqvist:2015sua} and thermalization takes place.
We consider the case where growth of $\phi$ field is not rapidly enough to relax to $\phi_M$ which is realized for $\xi \sim \mathcal{O}(1)$.
(See Ref.~\cite{Dimopoulos:2018wfg} for the case $\xi \gg1$ where $\phi$ soon settles down to its minimum after inflation.)

The effect of spinodal instability may be captured by Hartree approximation~\cite{Cormier:1999ia,Albrecht:2014sea} in which growth of instability is approximated by a Gaussian form. 
Under this approximation, one obtains $\phi^3\simeq 3\langle \phi^2\rangle \phi$ and $\phi^4 \simeq  3\langle \phi^2\rangle^2 $ where the coefficient is determined by Wick's theorem.
Here, $\langle \phi^2\rangle$ is defined by 
\begin{equation}
\begin{split}
	&\langle \phi^2 ({\bm x},t)\rangle \equiv \int \dfrac{\mathrm{d}k}{k}\mathcal{P}_\phi(k,t),\\
	&\phi({\bm x},t) = \int \frac{\mathrm{d}^3 k}{(2\pi)^{3/2}} \phi ({\bm k},t)e^{i{\bm k}\cdot {\bm x}},~\langle \phi ({\bm k},t)\phi^*({\bm k'},t)\rangle = \frac{2\pi^2}{k^3}\delta ({\bm k}-{\bm k'})\mathcal{P}_{\phi}(k,t) .
\end{split}
\end{equation}
The energy density of the SM Higgs field can be decomposed as 
\begin{equation}
\begin{split}
&\rho_{\rm Higgs}\equiv \rho_{\rm kin}+\rho_{\rm grad}+\rho_V,\\
&\rho_{\rm kin} = \frac{1}{2}\langle \dot{\phi}^2\rangle,~\rho_{\rm grad}=\frac{1}{2}\langle (\nabla \phi)^2\rangle,~\rho_V =  \frac
{3}{4}\lambda_{\rm SM}\langle \phi^2\rangle^2
 \label{eq:energy density of the SM Higgs}.
 \end{split}
\end{equation}
where $\rho_{\rm kin},~\rho_{\rm grad}$ and $\rho_V$ are the kinetic energy density, the gradient energy density and the potential energy density, respectively.

Introducing a rescaled field variable $\chi (\eta)\equiv a(\eta)\phi(\eta)$ and working in Fourier space, the equation of motion Eq.~\eqref{eq:equation of motion} can be written as 
\begin{align}
	\frac{d^2\chi_k (\eta)}{d\eta^2}  + \left[k^2 +3\lambda_{\rm SM}\langle \chi^2\rangle  +a^2(\eta)\left(\xi -\frac{1}{6}\right)R \right] \chi_k (\eta)=0. \label{eq:equation of motion of the SM Higgs field}
\end{align}
Here, $\chi_k$ is the mode function of $\chi$ and $\langle \chi^2\rangle $ is explicitly given by 
\begin{align}
	\langle \chi^2 \rangle  = \int \frac{dk}{k}P_{\chi},~P_\chi =\frac{k^3}{2\pi^2}|\chi_k (\eta)|^2. \label{eq:power specturm}
\end{align}
As for the boundary condition, we take the adiabatic vacuum in remote past:
\begin{align}
	\chi_k(\eta) = \sqrt{\frac{1}{2k}}e^{-ik\eta},~(k\eta\to-\infty). \label{eq:boundary conditions}
\end{align}
We numerically solve the equation of motion Eq.~\eqref{eq:equation of motion of the SM Higgs field} under the boundary condition Eq.~\eqref{eq:boundary conditions} using fourth order Runge-Kutta method from $H_{\rm inf}\eta=-10$.
In the numerical calculation, $a(\eta)$ is approximated by Eq.~\eqref{eq:scale factor}.
The mean field value of the SM Higgs field $\langle \chi^2\rangle$ is evaluated at each time step.
Integration with respect to momentum in Eq.~\eqref{eq:power specturm} is UV divergent, and hence, renormalization is required.
$\chi_k$ with modes satisfying $k\lesssim k_M \equiv -a^2(\eta_M)(\xi -1/6)R(\eta_M)$ experience spinodal instabilities where $\eta_M$ is the conformal time when $a^2(\eta)R$ takes minimum negative value, and thus, we simply introduce momentum cut-off $k_M$ to regularize the UV divergence.
The energy density of $\phi$ field is then evaluated by Eq.~\eqref{eq:energy density of the SM Higgs}.

\bibliographystyle{JHEP}
\bibliography{FIMP}

\end{document}